
\magnification=1200
\hsize=16.0truecm
\vsize=24.0truecm
\baselineskip=13pt
\pageno=0

\def\S{${\tilde \sigma}_{\c}$}
\def\s{\tilde{\sigma}_{\c}}
\def\J{$J/\psi$}
\def\j{J/\psi}
\def\P{$\psi'$}

\def\U{$\Upsilon$}
\def\u{\Upsilon}
\def\c{c{\bar c}}
\def\b{b{\bar b}}

\def\lsim{\raise0.3ex\hbox{$<$\kern-0.75em\raise-1.1ex\hbox{$\sim$}}}
\def\gsim{\raise0.3ex\hbox{$>$\kern-0.75em\raise-1.1ex\hbox{$\sim$}}}

\newcount\REFERENCENUMBER\REFERENCENUMBER=0
\def\REF#1{\expandafter\ifx\csname RF#1\endcsname\relax
               \global\advance\REFERENCENUMBER by 1
               \expandafter\xdef\csname RF#1\endcsname
                   {\the\REFERENCENUMBER}\fi}
\def\reftag#1{\expandafter\ifx\csname RF#1\endcsname\relax
               \global\advance\REFERENCENUMBER by 1
               \expandafter\xdef\csname RF#1\endcsname
                      {\the\REFERENCENUMBER}\fi
             \csname RF#1\endcsname\relax}
\def\ref#1{\expandafter\ifx\csname RF#1\endcsname\relax
               \global\advance\REFERENCENUMBER by 1
               \expandafter\xdef\csname RF#1\endcsname
                      {\the\REFERENCENUMBER}\fi
             [\csname RF#1\endcsname]\relax}
\def\refto#1#2{\expandafter\ifx\csname RF#1\endcsname\relax
               \global\advance\REFERENCENUMBER by 1
               \expandafter\xdef\csname RF#1\endcsname
                      {\the\REFERENCENUMBER}\fi
           \expandafter\ifx\csname RF#2\endcsname\relax
               \global\advance\REFERENCENUMBER by 1
               \expandafter\xdef\csname RF#2\endcsname
                      {\the\REFERENCENUMBER}\fi
             [\csname RF#1\endcsname--\csname RF#2\endcsname]\relax}
\def\refs#1#2{\expandafter\ifx\csname RF#1\endcsname\relax
               \global\advance\REFERENCENUMBER by 1
               \expandafter\xdef\csname RF#1\endcsname
                      {\the\REFERENCENUMBER}\fi
           \expandafter\ifx\csname RF#2\endcsname\relax
               \global\advance\REFERENCENUMBER by 1
               \expandafter\xdef\csname RF#2\endcsname
                      {\the\REFERENCENUMBER}\fi
            [\csname RF#1\endcsname,\csname RF#2\endcsname]\relax}
\def\refss#1#2#3{\expandafter\ifx\csname RF#1\endcsname\relax
               \global\advance\REFERENCENUMBER by 1
               \expandafter\xdef\csname RF#1\endcsname
                      {\the\REFERENCENUMBER}\fi
           \expandafter\ifx\csname RF#2\endcsname\relax
               \global\advance\REFERENCENUMBER by 1
               \expandafter\xdef\csname RF#2\endcsname
                      {\the\REFERENCENUMBER}\fi
           \expandafter\ifx\csname RF#3\endcsname\relax
               \global\advance\REFERENCENUMBER by 1
               \expandafter\xdef\csname RF#3\endcsname
                      {\the\REFERENCENUMBER}\fi
[\csname RF#1\endcsname,\csname RF#2\endcsname,\csname
RF#3\endcsname]\relax}
\def\refand#1#2{\expandafter\ifx\csname RF#1\endcsname\relax
               \global\advance\REFERENCENUMBER by 1
               \expandafter\xdef\csname RF#1\endcsname
                      {\the\REFERENCENUMBER}\fi
           \expandafter\ifx\csname RF#2\endcsname\relax
               \global\advance\REFERENCENUMBER by 1
               \expandafter\xdef\csname RF#2\endcsname
                      {\the\REFERENCENUMBER}\fi
            [\csname RF#1\endcsname,\csname RF#2\endcsname]\relax}
\def\Ref#1{\expandafter\ifx\csname RF#1\endcsname\relax
               \global\advance\REFERENCENUMBER by 1
               \expandafter\xdef\csname RF#1\endcsname
                      {\the\REFERENCENUMBER}\fi
             [\csname RF#1\endcsname]\relax}
\def\Refto#1#2{\expandafter\ifx\csname RF#1\endcsname\relax
               \global\advance\REFERENCENUMBER by 1
               \expandafter\xdef\csname RF#1\endcsname
                      {\the\REFERENCENUMBER}\fi
           \expandafter\ifx\csname RF#2\endcsname\relax
               \global\advance\REFERENCENUMBER by 1
               \expandafter\xdef\csname RF#2\endcsname
                      {\the\REFERENCENUMBER}\fi
            [\csname RF#1\endcsname--\csname RF#2\endcsname]\relax}
\def\Refand#1#2{\expandafter\ifx\csname RF#1\endcsname\relax
               \global\advance\REFERENCENUMBER by 1
               \expandafter\xdef\csname RF#1\endcsname
                      {\the\REFERENCENUMBER}\fi
           \expandafter\ifx\csname RF#2\endcsname\relax
               \global\advance\REFERENCENUMBER by 1
               \expandafter\xdef\csname RF#2\endcsname
                      {\the\REFERENCENUMBER}\fi
        [\csname RF#1\endcsname,\csname RF#2\endcsname]\relax}
\def\refadd#1{\expandafter\ifx\csname RF#1\endcsname\relax
               \global\advance\REFERENCENUMBER by 1
               \expandafter\xdef\csname RF#1\endcsname
                      {\the\REFERENCENUMBER}\fi \relax}

%

\def\NP{{ Nucl.\ Phys.\ }}
\def\PL{{ Phys.\ Lett.\ }}
\def\PR{{ Phys.\ Rev.\ }}

\def\PRL{{ Phys.\ Rev.\ Lett.\ }}

\def\ZP{{ Z.\ Phys.\ }}

\hfill CERN-TH.7526/94 \par
\hfill BI-TP 63/94
\vskip 3 truecm
\centerline{\bf QUARKONIUM PRODUCTION IN HADRONIC COLLISIONS}
\vskip 1truecm
\centerline{R. Gavai$^1$, D. Kharzeev$^{2,3}$, H. Satz$^{2,3}$}
\medskip
\centerline{G. A. Schuler$^2$, K. Sridhar$^2$, R. Vogt$^4$}
\vskip 3 truecm
\centerline{\bf Abstract:}
\medskip
We summarize the theoretical description of charmonium and
bottonium production in hadronic collisions and compare it to
the available data from hadron-nucleon interactions.
With the parameters of the theory established by these data,
we obtain predictions for quarkonium production
at RHIC and LHC energies.
\vskip 2 truecm
\centerline{To appear in}
\par
\centerline{{\sl Hard Processes in Hadronic Interactions}}
\par
\centerline{H. Satz and X.-N. Wang (Editors)}
\vfill
\hrule
{}~~~\par\noindent
1 Tata Institute of Fundamental Research, Bombay 400 005, India
\par\noindent
2 Theory Division, CERN, CH-1211 Geneva 23, Switzerland \par\noindent
3 Fakult\"at f\"ur Physik, Universit\"at Bielefeld, D-33501 Bielefeld,
Germany \par\noindent
4 Nuclear Science Division, LBL, University of California, Berkeley CA
94720, USA
\par\noindent
{}~~~\par\noindent
CERN-TH.7526/94 \par\noindent
BI-TP 63/94\par\noindent
December 1994 \par
\eject
The production of quarkonium states below the open charm/bottom
thresholds presents a particular challenge to QCD. Because of
the relatively large quark masses,
$\c$ and $\b$ production should be perturbatively calculable.
However, the subsequent transition from the
predominantly colour octet $Q{\bar Q}$ pairs to physical
quarkonium states can introduce nonperturbative aspects.
These may lead to some model-dependence, requiring cross
checks with as much data as possible.
\par
\refadd{Einhorn}
\refadd{Fritzsch}
\refadd{Reya}
\refadd{Sivers}
\refadd{Schuler}
A generalisation of the colour evaporation model
\refto{Einhorn}{Schuler} provides a unified approach to
the production of the different quarkonium states below the open
charm/bottom thresholds. As a specific example, we consider
charmonium production, although all arguments apply
to bottonium production as well. Parton-parton interactions
lead to the production of $\c$ pairs, as shown in Fig.\ 1.
We calculate the total ``hidden" charm cross section, \S, by
integrating over the $\c$ pair mass from $2m_c$ to
$2m_D$. At high energy, the dominant production mechanism is gluon
fusion (Fig.\ 1a), so that
$$
\s(s) = \int_{4m_c^2}^{4m_D^2} d\hat s \int dx_1 dx_2~g(x_1)
{}~g(x_2)~\sigma(\hat s)~\delta(\hat s-x_1x_2s), \eqno(1)
$$
with $g(x)$ denoting the gluon density and $\sigma$
the $gg \to \c$ cross section. In pion-nucleon collisions,
there are also significant quark-antiquark contributions (Fig.\ 1b),
which become dominant at low energies. Subsequently,
the $\c$ pair neutralizes its colour by interaction
with the collision-induced colour field (``colour evaporation").
During this process, the $c$ and the $\bar c$ either combine with
light quarks to produce charmed mesons, or they bind
with each other to form a charmonium state. More than half of
the subthreshold cross section \S~in fact goes into
open charm production (assuming $m_c~\lsim~1.5$ GeV);
the additional energy needed to produce charmed hadrons
is obtained (in general nonperturbatively) from the
colour field in the interaction region.
The yield of all charmonium states below the $D{\bar D}$ threshold is
thus only a part of the total sub-threshold cross section: in this
aspect the model we consider
is a generalisation of the original colour evaporation model
\refto{Einhorn}{Schuler}, which neglected the contribution of \S~to
open charm production. Using duality arguments, it equated \S~to the
sum over the charmonium states below the $D{\bar D}$ threshold.
\par
Neither the division of \S~into open charm and charmonia nor the
relative charmonium production rates are specified by the generalised
colour evaporation model. Hence its essential
prediction is that the energy dependence of charmonium production is
that of $\s(s)$. As a consequence, the ratios of different charmonium
production cross sections are energy-independent.
In Fig. 2, we show the ratio of \J~production from the decay $\chi_c
\to \gamma~\j$ to the total \J~production rate \refs{Schuler}{Anton1}.
It provides a measure of the $\chi_c/(\j)$ rate and is seen to be
independent of incident energy for both pion and proton beams. In Fig.
3, we show the measured \P/(\J) ratio \refss{Schuler}{Anton2}{Lou};
it is also found to be independent of the incident energy,
as well as of the projectile (pion or proton) and target (from
protons to the heaviest nuclei \ref{Ronceux}). Moreover,
it is noteworthy that the ratio \P/(\J) measured at
high transverse momenta at the Tevatron \ref{Teva}
is quite compatible with the $p_T$-integrated fixed target and ISR data
(Fig. 4).
\par
\refadd{Ueno}
\refadd{Yoshi}
\refadd{Moreno}
\refadd{Papadim}
The available bottonium data \refto{Ueno}{Papadim} also
agree with constant production ratios, as seen in Fig.\ 5 for the ratios
\U'/\U~and \U''/\U up to
Tevatron energies.
\par
The present data thus support one essential prediction of the colour
evaporation model up to 1.8 TeV. We now check if it also correctly
reproduces the variation of the production cross sections with
incident energy in this region.
In Figs.\ 6 and 7 we show the energy dependence of \J~production in
$pN$ collisions,
$$
\sigma_{pN\to\j}(s) = f_{\j}^p~\s(s), \eqno(2)
$$
as obtained from the hidden charm cross section $\s$
calculated in next-to-leading order \ref{Ridolfi} and with
the normalisation $f_{\j}^p$ fixed empirically.
We have
used the MRS D-' \ref{MRS} and GRV HO \ref{GRV} parametrisations
of the nucleon parton distributions functions
\ref{P-B}. For the GRV set, we have used $m_c=1.3$ GeV, with both
renormalisation and factorisation scales fixed to $m_c$. In
the MRS calculation, $m_c$=1.2 GeV was used, with the
scales set at $2m_c$. These parameters
provide an adequate description of open charm production, although the
results tend to lie somewhat below the measured total $\c$ cross
sections \ref{Ramona}. In Figs.\ 6 and 7 we show only
the MRS D-' result; the GRV HO result differs by less than 5 \% in
this energy range.
The agreement with the data \ref{Schuler} over the entire
range is quite satisfactory, with the normalisation $f_{\j}^p=0.025$.
In Fig.\ 8 we find equally good agreement
for the energy dependence of \J~production with pion beams. However,
the fraction of \J~in the hidden charm cross section must
be slightly higher to reproduce the pion
data well, with $f_{\j}^{\pi}=
0.034$ for a good fit. This may well be due to
greater uncertainties in the pionic parton distribution functions.
We have also calculated the leading order cross section; the resulting
theoretical K-factor, $K \equiv \s^{NLO}/\s^{LO}$, remains between 2.0
and 2.5 over the currently measured energy range for both sets of
parton distribution functions and for pion and proton beams.
\par
The fraction of \S~producing charmonium rather than open charm
is thus about 10\%. This is in
accord with open charm calculations, which show \ref{Ramona} that
much of the total cross section comes from
subthreshold $\c$ initial states which acquire the necessary energy for
$D{\bar D}$ formation from the interaction colour field. To
illustrate this, Fig.\ 9 shows the fraction of the total
open charm cross section with
$2~m_c \leq M \leq 2~M_D$, where $M\equiv \sqrt {\hat s}$.
It remains quite large even at very high incident energies.
\par
We further compare the longitudinal momentum dependence of charmonium
production with recent experimental results.
In fig.\ 10 we compare data with our
calculations for the $x_F$ dependence of \J~ production at several
energies and for $\pi-p,~{\bar p}-p$ and $p-p$ collisions \ref{x-F}.
Since there is a spread of integrated cross section values
around the average $\s$, as seen in Figs. 6 - 8, we have normalised
the calculated $x_F$ distribution to the integrated
experimental one. We conclude that the $x_F$ distributions are also
consistent with the colour evaporation model.
\par
Next we comment briefly on the transverse momentum distributions.
We are
interested in low $p_T$ charmonium production, for which the
model
provides essentially no prediction. There is the intrinsic
transverse momentum of the initial partons, the
intrinsic momentum fluctuations of the colour field which neutralises
the colour of the $\c$ system in the evaporation process,
and at larger $p_T$ higher order perturbative terms. Since there is no
way to separate these different contributions in the low $p_T$ region,
the model has no predictive power.
\par
The colour evaporation model thus reproduces correctly both the energy
dependence and the $x_F$ distributions of charmonium production,
up to an open normalisation constant for
each charmonium state, which can be fixed empirically by
data. Once this is done, integrated and differential
cross sections can be predicted for RHIC and LHC energies. From
the fits to the data shown in Figs.\ 6 and 7 we obtain
$$
\left({d\sigma_{pN\to \j}\over dy}\right)_{y=0}
= 2.5~\times 10^{-2}~\left( {d\s^{NLO}\over dy} \right)_{y=0}
\eqno(3)
$$
for \J~production.
In Fig.\ 11 we show the resulting $(d\sigma_{pN \to \j}/dy)_{y=0}$ as
function of the center of mass energy, $\sqrt s$, and in Fig.\ 12 we give
the rapidity distributions at RHIC and LHC energies.
The cross sections are listed in Table 1.
\par
Before commenting on our predictions, we first
repeat the analysis for \U~production. Because the data
generally give the sum
of \U, \U' and \U'' production, the branching ratios cannot simply be
removed. Therefore we show in Fig.\ 13 the measured cross section
for the sum of the three \U~states in the dilepton decay
channel, denoted by $B(d\sigma/dy)_{y=0}$. We see that
$$
B\left({d\sigma_{pN\to \u}\over dy}\right)_{y=0}
= 1.6~\times 10^{-3}~\left( {d\sigma_{b{\bar b}}^{NLO}\over dy}
\right)_{y=0}  \eqno(4)
$$
gives a good description
of the data up to and including ISR results.
The results are also calculated using the
MRS D-' and GRV HO parton distribution functions, with
$m_b=4.75$ GeV and the scales equal to $m_b$.
Assuming the bulk of the cross section to be from \U(1S) production,
and using the corresponding branching ratio, we estimate from eq.\ (4)
that about 7\% of the sub-threshold $\b$ cross
section leads to $\Upsilon$ production.
\par
Using the normalisaton determined in Eq.\ (4), we obtain the cross
section for high energy \U~production;
the results are shown in Figs. 14 and
15 and in Table 2. The recent high energy data
from UA1 \ref{Eggert} and CDF \ref{Papadim}
agree very well with the predicted energy dependence,
as seen in Fig.\ 13,
giving strong support to
the ``new" parton distribution functions based on HERA data \ref{HERA}.
They also give us considerable confidence in the extrapolation to
LHC energy.
\par
We now comment on some features of our predictions.
The two parton distribution functions,
MRS D-' and GRV HO, provide fully compatible
results in the measured energy range. The
\U~predictions agree with data even up to energies close to the LHC
range. The MRS D-' \J~cross section is
about twice as large as the GRV HO prediction at LHC energies.
This is because the MRS distributions require larger factorisation
scales than the GRV distributions. Both parton distribution functions,
with their chosen scales, also give acceptable fits to the measured
open charm production cross sections (see \ref{Ramona} for more
details).
The difference thus gives some indication of the uncertainty of the
\J~prediction. The \U results agree over the entire energy range,
since $m_b=4.75$ GeV was used as the scale in both sets.
\par
At LHC energies, both the \J~and \U~rapidity distributions
remain rather constant out to $y\simeq 4$, using the MRS parton
distributions. The GRV HO results show an even
wider plateau. In either case, there is a
large window for forward detection at high energies.
At RHIC energies, the \J~distributions are not as broad, with a
forward plateau of 2 - 3 units for the MRS set,
while the GRV distributions are somewhat
narrower. The \U~rapidity distributions at RHIC energies are quite
similar for both sets.
\par
\refadd{Craigie}
\refadd{O/Ramona}
\refadd{Gatlin}
Finally we note that the cross sections calculated with the
recent parton distribution functions
are considerably higher, typically by about a
factor 20, than those given by an earlier empirical parametrisation,
$\sim {\rm exp}(-15 M/\sqrt s)$,
labelled CR in Figs.\ 11 and 14 \refto{Craigie}{Gatlin} .
This increase, confirmed by new high energy \U~data
(Fig.\ 14), is mainly due to the increase
in the gluon distribution functions at small $x$,
as suggested by data
from HERA \ref{HERA}.
\par
The colour evaporation model addresses the common energy
behaviour of the different
quarkonium states. To determine their relative production
rates, the colour evaporation process has to be specified in
more detail. Let us consider one example of this. Assume that the
initial colour octet state first neutralises its colour by
interaction with the surrounding colour field, producing a colour
singlet $\c$ state. The relative weights of \J~and \P~production can
then be expressed \ref{Schuler}
in terms of the corresponding masses and the
squared charmonium wave functions at the origin,
$$
{\sigma(\psi') \over \sigma(\psi)} = {R_{\psi'}^2(0) \over
R_{\psi}^2(0)}\left({M_{\psi}\over M_{\psi'}}\right)^5. \eqno(5)
$$
Here $\psi$ denotes the directly produced 1S $\c$ state, in contrast
to the experimentally observed \J, 40\% of which originates
from radiative $\chi_c$ decays (see Fig.\ 1). The wave functions at the
origin can in turn be related to the dilepton decay widths
$\Gamma_{ee} \sim (R^2(0) / M^2)$ \ref{Schuler}, giving
$$
{\sigma(\psi') \over \sigma(\psi)} = {\Gamma_{\psi'} \over
\Gamma_{\psi}}\left({M_{\psi}\over M_{\psi'}} \right)^3. \eqno(6)
$$
Inserting the measured masses and decay widths, we find
$$
{\sigma(\psi') \over \sigma(\psi)} \simeq 0.24~. \eqno(7)
$$
To compare this to the measured value of
$\sigma(\psi')/\sigma(\psi)$,
we have to remove the $\chi_c$ contributions from the
experimental ratio,
$$
{\sigma(\psi') \over \sigma(\psi)} = \left[{1\over
1-(\sigma_{\chi_c}/\sigma_{\j})}\right] \left[
{\sigma(\psi') \over \sigma(\j)}\right]_{\rm exp}. \eqno(8)
$$
With the experimental values $\sigma(\psi')/\sigma(\j)\simeq 0.14$
(see Figs.\ 2 and 3) and
$(\sigma_{\chi_c}/\sigma_{\j})\simeq 0.4$ (see Fig.\
1), this yields $\sigma(\psi')/ \sigma(\psi)\simeq 0.23$, in good
agreement with the theoretical result (7). We thus find that the
projection of the colour singlet $\c$ state onto \J~and \P~correctly
describes their production ratios at all energies and transverse
momenta.
\par
The predictions for direct bottonium production ratios corresponding to
eq.\ (7) are
$$
{\sigma(\Upsilon') \over \sigma(\Upsilon)} \simeq 0.36~~;~~
{\sigma(\Upsilon'') \over \sigma(\Upsilon)} \simeq 0.27~. \eqno(9)
$$
Since the contributions from indirect production through
radiative $\chi_b$ decay are not yet known and there is also
feeding from higher S-states, a quantitative comparison
is not possible here. Nevertheless, the predicted values differ
from the data (see Fig.\ 5) by less than 50 \%
and hence appear reasonable.
\par
So far, the most complete description of the colour evaporation process
is attempted in the colour singlet model
\ref{CS}, in which not only the $\c$ formation but also the subsequent
colour neutralisation is assumed to take place on a perturbative scale.
The resonance formation is then determined by the
appropriate wave functions with the right quantum numbers, as
above. As a result, the production cross section for each
charmonium state is completely determined to the order of
perturbation theory used. Some characteristic production diagrams in
lowest order are illustrated
in Fig.\ 16. As generally formulated, the scale of the strong coupling
constant in all perturbative diagrams is determined by the mass of the
heavy quark.
\par
Such a perturbative description of colour neutralisation can be
justified only if {\sl all} momentum scales are sufficiently large.
However, as seen in Fig.\ 16, colour neutralisation for all but
$\eta_c$, $\chi_0$ and
and $\chi_2$ requires the emission or absorption of a ``third" gluon.
This restricts the possible applicability of the model to
production
at large transverse momentum. In the $p_T$-integration, the ``third"
gluon is soft
($k \sim \Lambda_{\rm QCD}$) in a significant part of phase space, and
hence the model becomes unreliable here even though the
integration is infrared finite.
It is therefore not surprising that the colour singlet
model leads to charmonium production ratios which disagree rather
strongly with experiment.
The quantum numbers of the $\chi_2$ allow partonic production at order
$\alpha_s^2$, while \J, $\chi_1$ and \P~production are all of order
$\alpha_s^3$. As a result, their production is much too strongly
suppressed in comparison to the $\chi_2$. Thus, while the model
predicts $\chi_2/(\j) \simeq 10$, the measured ratios are
below 2 \ref{Anton1}. The inclusion of certain relativistic corrections
can somewhat reduce this discrepancy \ref{Schuler}. --
Similar arguments hold for \U~production, although the soft part of
the $p_T$-integration is relatively smaller, so that here the predictions
may be closer to the data.

\par
For the validity of a perturbative treatment,
the ``third" gluon has to be hard enough to
resolve the $\c$ into individual quarks. Hence its momentum must be
higher than $1/r_{\j} \simeq 1/(0.2~{\rm fm}) \simeq 1$ GeV. Below this
limit, it is not clear how colour neutralisation is achieved;
presumably nonperturbative
interactions of the colour octet $\c$ with the gluon condensate play a
considerable role. Some additional contributions can
perhaps also be obtained by summing classes of perturbative contributions
\ref{Neubert}. However, as long as the additional
interactions cannot be determined quantitatively, the prediction of the
$p_T$-integrated production ratios of the different charmonium states
is not possible. It is not known if this is also true for
bottonium production, or if here the role of soft processes has
become sufficiently reduced to allow a fully perturbative treatment.
\par
\vfill\eject
\centerline{\bf References:}
\bigskip
\item{\reftag{Einhorn})}{M. B. Einhorn and S. D. Ellis, \PR D12 (1975)
2007.}
\item{\reftag{Fritzsch})}{H. Fritzsch, \PL 67B (1977) 217.}
\item{\reftag{Reya})}
{M. Gl\"uck, J. F. Owens and E. Reya, \PR D17 (1978) 2324.}
\item{\reftag{Sivers})}
{J. Babcock, D. Sivers and S. Wolfram, \PR D18 (1978) 162.}
\item{\reftag{Schuler})}
{For a recent review, see G. A. Schuler, ``Quarkonium Production and
Decays", CERN Preprint CERN-TH.7170/94, February 1994.}
\item{\reftag{Anton1})}{L. Antoniazzi et al., \PRL 70 (1993) 383.}
\item{\reftag{Anton2})}{L. Antoniazzi et al., \PR D46 (1992) 4828.}
\item{\reftag{Lou})}{C. Lourenco, ``\J,\P~and Dimuon Production in
$p-A$ and $S-U$ Collisions at 200 GeV/Nucleon", Dissertation,
Technical University of Lisbon, Portugal.}
\item{\reftag{Ronceux})}{B. Ronceux, \NP A566 (1994) 371c.}
\item{\reftag{Teva})}{The CDF Collaboration, ``$\j,~\psi' \to
\mu^+\mu^-$ and $B \to \j,~\psi'$ Cross Sections", Fermilab Preprint
Fermilab-Conf-94/136-E, May 1994 (Contribution to the International
Conference on High Energy Physics, Glasgow, Scotland 1994);\hfill\par
E. Braaten et al., \PL B333 (1994) 548.}
\item{\reftag{Ueno})}{K. Ueno et al., \PRL 42 (1979) 486.}
\item{\reftag{Yoshi})}{T. Yoshida et al., \PR D 39 (1989) 3516.}
\item{\reftag{Moreno})}{G. Moreno et al., \PR D 43 (1991) 2815.}
\item{\reftag{Papadim})}{V. Papadimitriou (CDF), ``\U~Production at
CDF", Fermilab Preprint Fermilab-Conf-94-221-E, August 1994.}
\item{\reftag{Ridolfi})}{M. L. Mangano, P. Nason and G. Ridolfi, \NP
B405 (1993) 507.}
\item{\reftag{MRS})}{A. D. Martin, R. G. Roberts and W. J. Stirling, \PL
B 306 (1993) 145, \hfill\break and ``Structure Functions and Parton
Distributions, in this volume.}
\item{\reftag{GRV})}{M. Gl\"uck, E. Reya and A. Vogt, \ZP C53 (1993)
127.}
\item{\reftag{P-B})}{H. Plothow-Besch, Comp. Phys. Comm. 75 (1993) 396,
\hfill\break and ``PDFLIB", in this volume.}
\item{\reftag{Ramona})}{P. L. McGaughey et al., ``Heavy Quark
Production in $pp$ Collisons", in this volume.}
\item{\reftag{x-F})}{C. Akerlof et al., \PR D48 (1993) 5067;\hfill\break
L. Antoniazzi et al., \PR D46 (1992) 4828;\hfill\break
M. S. Kowitt et al., \PRL 72 (1994) 1318;\hfill\break
C. Biino et al., \PRL 58 (1987) 2523;\hfill\break
V. Abramov et al., ``Properties of \J~production in $\pi^--Be$ and
$p-Be$ Collisions at 530 GeV/c", Fermilab Preprint
FERMILAB-PUB-91/62-E.}
\item{\reftag{Eggert})}{K. Eggert and A. Morsch (UA1), private
communication.}
\item{\reftag{HERA})}{M. Derrick et al. (Zeus), \PL B 316 (1993) 412;
\hfill\break
I. Abt et al. (H1), \NP B 407 (1993) 515.}
\item{\reftag{Craigie})}{N. Craigie, Phys. Rep. 47 (1978) 1.}
\item{\reftag{O/Ramona})}{R.Vogt, Atomic Data and Nuclear Data Tables 50
(1992) 343.}
\item{\reftag{Gatlin})}{H. Satz, \NP A544 (1992) 371c.}
\vfill\eject
\item{\reftag{CS})}{C. H. Chang, \NP B 172 (1980) 425;\hfill\break
E. L. Berger and D. Jones, \PR D 23 (1981) 1521;\hfill\break
R. Baier and R. R\"uckl, \PL B 102 (1981) 364 and \ZP C 19 (1983) 251.}
\item{\reftag{Neubert})}{M. Beneke and V. M. Braun, ``Naive
Non-Abelianisation and Resummation of Fermion Bubble Chains",
DESY Preprint DESY 94-200, November 1994; \hfill\break
M. Neubert, ``Scale Setting in QCD and the Momentum Flow in Feynman
Diagrams", CERN Preprint CERN-TH.7487/94, December 1994.}
\vfill\eject

{}~~~\vskip 1 truecm
\centerline{\bf Table 1:  $J/\psi$ Production}
\bigskip
$$
{\offinterlineskip \tabskip=0pt
\vbox{
\halign to 0.9\hsize
{\strut
\vrule width0.8pt\quad#
\tabskip=0pt plus100pt
& # \quad
&\vrule#&
&\quad \hfil # \quad
&\vrule#
&\quad \hfil # \quad
\tabskip=0pt
&\vrule width0.8pt#
\cr
\noalign{\hrule}\noalign{\hrule}
&~~~~~~ &&~~~~~~~~~~~~~~~~~~~~~ &&~~~~~&\cr
&$\sqrt{s}$~~~[GeV] &&$(d\sigma/dy)_{y=0}^{\rm MRS}$~~~[$\mu$b] &&
$(d\sigma/dy)_{y=0}^{\rm GRV}$~~~[$\mu$b]  &\cr
&~~~~~~ &&~~~~~~~~~~~~~~~~~~~~~ &&~~~~~&\cr
\noalign{\hrule}
&~~    &&~~                      &&~                       &\cr
&\hfill   20 &&~~~$6.2 \times 10^{-2}$ &&~~~$5.8 \times 10^{-2}$ &\cr
&\hfill   40 &&~~~$1.6 \times 10^{-1}$ &&~~~$1.5 \times 10^{-1}$ &\cr
&\hfill   60 &&~~~$2.5 \times 10^{-1}$ &&~~~$2.4 \times 10^{-1}$ &\cr
&\hfill  100 &&~~~$3.5 \times 10^{-1}$ &&~~~$3.4 \times 10^{-1}$ &\cr
&\hfill  200 &&~~~$6.3 \times 10^{-1}$ &&~~~$5.9 \times 10^{-1}$ &\cr
&\hfill  500 &&~~~$1.5 \times 10^{+0}$ &&~~~$1.2 \times 10^{+0}$ &\cr
&\hfill 1000 &&~~~$3.2 \times 10^{+0}$ &&~~~$2.5 \times 10^{+0}$ &\cr
&\hfill 5500 &&~~~$1.6 \times 10^{+1}$ &&~~~$5.9 \times 10^{+0}$ &\cr
&\hfill14000 &&~~~$4.1 \times 10^{+1}$ &&~~~$1.1 \times 10^{+1}$ &\cr
&~~  &&~~                      &&~                       &\cr
\noalign{\hrule}}
}}
$$
\vskip 2 truecm
\centerline{\bf Table 2:  $(\Upsilon+\Upsilon'+\Upsilon'')$
Production}
\bigskip
$$
{\offinterlineskip \tabskip=0pt
\vbox{
\halign to 0.9\hsize
{\strut
\vrule width0.8pt\quad#
\tabskip=0pt plus100pt
& # \quad
&\vrule#&
&\quad \hfil # \quad
&\vrule#
&\quad \hfil # \quad
\tabskip=0pt
&\vrule width0.8pt#
\cr
\noalign{\hrule}\noalign{\hrule}
&~~~~~~ &&~~~~~~~~~~~~~~~~~~~~~ &&~~~~~&\cr
&$\sqrt{s}$~~~[GeV] &&$(Bd\sigma/dy)_{y=0}^{\rm MRS}$~~~[pb] &&
$(Bd\sigma/dy)_{y=0}^{\rm GRV}$~~~[pb]  &\cr
&~~~~~~ &&~~~~~~~~~~~~~~~~~~~~~ &&~~~~~&\cr
\noalign{\hrule}
&~~    &&~~                      &&~                       &\cr
&\hfill   15 &&~~~$3.1 \times 10^{-4}$ &&~~~$2.5 \times 10^{-4}$ &\cr
&\hfill   30 &&~~~$9.7 \times 10^{-1}$ &&~~~$9.7 \times 10^{-1}$ &\cr
&\hfill   60 &&~~~$1.2 \times 10^{+1}$ &&~~~$1.2 \times 10^{+1}$ &\cr
&\hfill  100 &&~~~$3.4 \times 10^{+1}$ &&~~~$3.7 \times 10^{+1}$ &\cr
&\hfill  200 &&~~~$8.6 \times 10^{+1}$ &&~~~$1.0 \times 10^{+2}$ &\cr
&\hfill  500 &&~~~$2.5 \times 10^{+2}$ &&~~~$3.4 \times 10^{+2}$ &\cr
&\hfill 1000 &&~~~$5.5 \times 10^{+2}$ &&~~~$8.8 \times 10^{+2}$ &\cr
&\hfill 5500 &&~~~$3.0 \times 10^{+3}$ &&~~~$3.6 \times 10^{+3}$ &\cr
&\hfill14000 &&~~~$7.8 \times 10^{+3}$ &&~~~$7.7 \times 10^{+3}$ &\cr
&~~  &&~~                      &&~                       &\cr
\noalign{\hrule}}
}}
$$
\vfill\eject

\centerline{\bf Figure Captions}
\bigskip
\parindent=0pt
Fig.\ 1: Lowest order $\c$ production through gluon fusion (a) and
quark-antiquark annihilation (b).
\medskip
Fig.\ 2: The ratio of $(\chi_1 + \chi_2)\to \j$ to total \J~production,
as a function of the center of mass
energy, $\sqrt s$, by proton
(open symbols) and pion beams (solid symbols) \ref{Anton1}.
\medskip
Fig.\ 3a: The ratio of $\psi'$ to \J~production as a function of
the center of mass energy, $\sqrt s$, on proton
(circles) and nuclear targets (squares) \refss{Schuler}{Anton2}{Lou}.
The average value is $0.14~\pm~0.03$.
\medskip
Fig.\ 3b: The ratio of $\psi'$ to \J~production by proton beams as a
function of the atomic mass number $A$ for data in the energy range $20
\leq \sqrt s \leq 40$ GeV \ref{Ronceux}. The average value is
$0.14~\pm~0.01$.
\medskip
Fig.\ 4a: The ratio of $\psi'$ to \J~production as a function of
transverse momentum \ref{Teva}; the shaded strip shows the
average value of Fig.\ 3.
\medskip
Fig.\ 4b: The ratio of $\psi'$ to \J~production as a function of
center of mass energy, $\sqrt s$. The fixed target and ISR data are
integrated over the low $p_T$ region, while the CDF point is the average
over $5 \leq p_T \leq 15$ GeV.
\medskip
Fig.\ 5: The ratios of \U' and \U'' to \U~production as a function of the
center of mass energy, $\sqrt s$ \refto{Ueno}{Papadim}.
The average values are 0.53$\pm$0.13 and
0.17$\pm$0.06, respectively.
\medskip
Fig.\ 6: The differential \J~production cross section
$(d\sigma_{\j}^{pN}/dy)=
2.5\times10^{-2}~(d\s^{pN}/dy)$ at $y=0$, calculated with
MRS D-' parton distributions, compared to data \ref{Schuler}.
\medskip
Fig.\ 7: The \J~production cross section $\sigma_{\j}^{pN}=2.5 \times
10^{-2}~\s^{pN}$ for $x_F>0$,
calculated with MRS D-' parton distributions, compared
to data \ref{Schuler}.
\medskip
Fig.\ 8: The \J~production cross section $\sigma_{\j}^{\pi
N}=3.4\times 10^{-2}~\s^{\pi N}$ for $x_F>0$,
calculated with
MRS D-'/SMRS P2 parton distributions, compared to data \ref{Schuler}.
\medskip
Fig.\ 9: The fraction of the total open charm cross section due to
the ``hidden" charm mass interval $[2m_c,2m_D]$.
\medskip
Fig.\ 10a: The \J~longitudinal momentum distributions compared to
${\bar p}N$ and $pN$ data \ref{x-F},
 with $x_F=p_L(\j)/p_{max}(\j)$;
the MRS results are denoted by a solid, the GRV by a dashed line.
\medskip
Fig.\ 10b: The \J~longitudinal momentum distributions compared to
$\pi N$ data \ref{x-F}, with $x_F=p_L(\j)/p_{max}(\j)$.
the MRS results are denoted by a solid, the GRV by a dashed line.
\medskip
Fig.\ 11: Energy dependence of $(d\sigma_{\j}^{pN}/dy)_{y=0}$
for \J~production, as obtained with MRS D-' and GRV HO
parton distributions.
\medskip
Fig.\ 12: Rapidity distributions for \J~production, calculated with MRS
D-' parton distributions at RHIC and LHC energies.
\medskip
Fig.\ 13: The differential $\Upsilon$ production cross section
$(d\sigma_{\u}^{pN}/dy)=1.8\times 10^{-3}
{}~(d\sigma_{b{\bar b}}^{pN}/dy)$ at $y=0$, calculated with
MRS D-' parton distributions, compared to data
\ref{Schuler}. The corresponding GRV HO predictions are very similar.
\medskip
Fig.\ 14: Energy dependence of $(d\sigma_{\u}^{pN}/dy)_{y=0}
$ for \U~production, with high energy data from \refs{Eggert}{Papadim};
the predictions of MRS D-' and GRV HO
essentially coincide. Also shown (CR) is the phenomenological fit of
\ref{Craigie}.
\medskip
Fig.\ 15: Rapidity distributions for \U~production calculated with MRS
D-' parton distributions at RHIC and LHC energies.
\medskip
Fig.\ 16: Lowest order contributions to
charmonium production in the colour singlet model.
\medskip
\vfill\eject\bye

\magnification=1200
\baselineskip 13pt
\nopagenumbers

\end
\magnification=1200
\hsize=16.0truecm
\vsize=24.0truecm
\baselineskip=13pt
\pageno=0

\def\S{${\tilde \sigma}_{\c}$}
\def\s{\tilde{\sigma}_{\c}}
\input macro-hs

{}~~~\vskip 1 truecm
\centerline{\bf Table 1:  $J/\psi$ Production}
\bigskip
$$
{\offinterlineskip \tabskip=0pt
\vbox{
\halign to 0.9\hsize
{\strut
\vrule width0.8pt\quad#
\tabskip=0pt plus100pt
& # \quad
&\vrule#&
&\quad \hfil # \quad
&\vrule#
&\quad \hfil # \quad
\tabskip=0pt
&\vrule width0.8pt#
\cr
\noalign{\hrule}\noalign{\hrule}
&~~~~~~ &&~~~~~~~~~~~~~~~~~~~~~ &&~~~~~&\cr
&$\sqrt{s}$~~~[GeV] &&$(d\sigma/dy)_{y=0}^{\rm MRS}$~~~[$\mu$b] &&
$(d\sigma/dy)_{y=0}^{\rm GRV}$~~~[$\mu$b]  &\cr
&~~~~~~ &&~~~~~~~~~~~~~~~~~~~~~ &&~~~~~&\cr
\noalign{\hrule}
&~~    &&~~                      &&~                       &\cr
&\hfill   20 &&~~~$6.2 \times 10^{-2}$ &&~~~$5.8 \times 10^{-2}$ &\cr
&\hfill   40 &&~~~$1.6 \times 10^{-1}$ &&~~~$1.5 \times 10^{-1}$ &\cr
&\hfill   60 &&~~~$2.5 \times 10^{-1}$ &&~~~$2.4 \times 10^{-1}$ &\cr
&\hfill  100 &&~~~$3.5 \times 10^{-1}$ &&~~~$3.4 \times 10^{-1}$ &\cr
&\hfill  200 &&~~~$6.3 \times 10^{-1}$ &&~~~$5.9 \times 10^{-1}$ &\cr
&\hfill  500 &&~~~$1.5 \times 10^{+0}$ &&~~~$1.2 \times 10^{+0}$ &\cr
&\hfill 1000 &&~~~$3.2 \times 10^{+0}$ &&~~~$2.5 \times 10^{+0}$ &\cr
&\hfill 5500 &&~~~$1.6 \times 10^{+1}$ &&~~~$5.9 \times 10^{+0}$ &\cr
&\hfill14000 &&~~~$4.1 \times 10^{+1}$ &&~~~$1.1 \times 10^{+1}$ &\cr
&~~  &&~~                      &&~                       &\cr
\noalign{\hrule}}
}}
$$
\vskip 2 truecm
\centerline{\bf Table 2:  $(\Upsilon+\Upsilon'+\Upsilon'')$
Production}
\bigskip
$$
{\offinterlineskip \tabskip=0pt
\vbox{
\halign to 0.9\hsize
{\strut
\vrule width0.8pt\quad#
\tabskip=0pt plus100pt
& # \quad
&\vrule#&
&\quad \hfil # \quad
&\vrule#
&\quad \hfil # \quad
\tabskip=0pt
&\vrule width0.8pt#
\cr
\noalign{\hrule}\noalign{\hrule}
&~~~~~~ &&~~~~~~~~~~~~~~~~~~~~~ &&~~~~~&\cr
&$\sqrt{s}$~~~[GeV] &&$(Bd\sigma/dy)_{y=0}^{\rm MRS}$~~~[pb] &&
$(Bd\sigma/dy)_{y=0}^{\rm GRV}$~~~[pb]  &\cr
&~~~~~~ &&~~~~~~~~~~~~~~~~~~~~~ &&~~~~~&\cr
\noalign{\hrule}
&~~    &&~~                      &&~                       &\cr
&\hfill   15 &&~~~$3.1 \times 10^{-4}$ &&~~~$2.5 \times 10^{-4}$ &\cr
&\hfill   30 &&~~~$9.7 \times 10^{-1}$ &&~~~$9.7 \times 10^{-1}$ &\cr
&\hfill   60 &&~~~$1.2 \times 10^{+1}$ &&~~~$1.2 \times 10^{+1}$ &\cr
&\hfill  100 &&~~~$3.4 \times 10^{+1}$ &&~~~$3.7 \times 10^{+1}$ &\cr
&\hfill  200 &&~~~$8.6 \times 10^{+1}$ &&~~~$1.0 \times 10^{+2}$ &\cr
&\hfill  500 &&~~~$2.5 \times 10^{+2}$ &&~~~$3.4 \times 10^{+2}$ &\cr
&\hfill 1000 &&~~~$5.5 \times 10^{+2}$ &&~~~$8.8 \times 10^{+2}$ &\cr
&\hfill 5500 &&~~~$3.0 \times 10^{+3}$ &&~~~$3.6 \times 10^{+3}$ &\cr
&\hfill14000 &&~~~$7.8 \times 10^{+3}$ &&~~~$7.7 \times 10^{+3}$ &\cr
&~~  &&~~                      &&~                       &\cr
\noalign{\hrule}}
}}
$$
\vfill\eject
\bye
\centerline{\bf Figure Captions}
\bigskip
\parindent=0pt
Fig.\ 1: Lowest order $\c$ production through gluon fusion (a) and
quark-antiquark annihilation (b).
\medskip
Fig.\ 2: The ratio of $(\chi_1 + \chi_2)\to \j$ to total \J~production,
as a function of the center of mass
energy, $\sqrt s$, by proton
(open symbols) and pion beams (solid symbols) \ref{Anton1}.
\medskip
Fig.\ 3a: The ratio of $\psi'$ to \J~production as a function of
the center of mass energy, $\sqrt s$, on proton
(circles) and nuclear targets (squares) \refss{Schuler}{Anton2}{Lou}.
The average value is $0.14~\pm~0.03$.
\medskip
Fig.\ 3b: The ratio of $\psi'$ to \J~production by proton beams as a
function of the atomic mass number $A$ for data in the energy range $20
\leq \sqrt s \leq 40$ GeV \ref{Ronceux}. The average value is
$0.14~\pm~0.01$.
\medskip
Fig.\ 4a: The ratio of $\psi'$ to \J~production as a function of
transverse momentum \ref{Teva}; the shaded strip shows the
average value of Fig.\ 3.
\medskip
Fig.\ 4b: The ratio of $\psi'$ to \J~production as a function of
center of mass energy, $\sqrt s$. The fixed target and ISR data are
integrated over the low $p_T$ region, while the CDF point is the average
over $5 \leq p_T \leq 15$ GeV.
\medskip
Fig.\ 5: The ratios of \U' and \U'' to \U~production as a function of the
center of mass energy, $\sqrt s$ \refto{Ueno}{Papadim}.
The average values are 0.53$\pm$0.13 and
0.17$\pm$0.06, respectively.
\medskip
Fig.\ 6: The differential \J~production cross section
$(d\sigma_{\j}^{pN}/dy)=
2.5\times10^{-2}~(d\s^{pN}/dy)$ at $y=0$, calculated with
MRS D-' parton distributions, compared to data \ref{Schuler}.
\medskip
Fig.\ 7: The \J~production cross section $\sigma_{\j}^{pN}=2.5 \times
10^{-2}~\s^{pN}$ for $x_F>0$,
calculated with MRS D-' parton distributions, compared
to data \ref{Schuler}.
\medskip
Fig.\ 8: The \J~production cross section $\sigma_{\j}^{\pi
N}=3.4\times 10^{-2}~\s^{\pi N}$ for $x_F>0$,
calculated with
MRS D-'/SMRS P2 parton distributions, compared to data \ref{Schuler}.
\medskip
Fig.\ 9: The fraction of the total open charm cross section due to
the ``hidden" charm mass interval $[2m_c,2m_D]$.
\medskip
Fig.\ 10a: The \J~longitudinal momentum distributions compared to
${\bar p}N$ and $pN$ data \ref{x-F},
 with $x_F=p_L(\j)/p_{max}(\j)$;
the MRS results are denoted by a solid, the GRV by a dashed line.
\medskip
Fig.\ 10b: The \J~longitudinal momentum distributions compared to
$\pi N$ data \ref{x-F}, with $x_F=p_L(\j)/p_{max}(\j)$.
the MRS results are denoted by a solid, the GRV by a dashed line.
\medskip
Fig.\ 11: Energy dependence of $(d\sigma_{\j}^{pN}/dy)_{y=0}$
for \J~production, as obtained with MRS D-' and GRV HO
parton distributions.
\medskip
Fig.\ 12: Rapidity distributions for \J~production, calculated with MRS
D-' parton distributions at RHIC and LHC energies.
\medskip
Fig.\ 13: The differential $\Upsilon$ production cross section
$(d\sigma_{\u}^{pN}/dy)=1.8\times 10^{-3}
{}~(d\sigma_{b{\bar b}}^{pN}/dy)$ at $y=0$, calculated with
MRS D-' parton distributions, compared to data
\ref{Schuler}. The corresponding GRV HO predictions are very similar.
\medskip
Fig.\ 14: Energy dependence of $(d\sigma_{\u}^{pN}/dy)_{y=0}
$ for \U~production, with high energy data from \refs{Eggert}{Papadim};
the predictions of MRS D-' and GRV HO
essentially coincide. Also shown (CR) is the phenomenological fit of
\ref{Craigie}.
\medskip
Fig.\ 15: Rapidity distributions for \U~production calculated with MRS
D-' parton distributions at RHIC and LHC energies.
\medskip
Fig.\ 16: Lowest order contributions to
charmonium production in the colour singlet model.
\medskip
\vfill\eject\bye

\magnification=1200
\baselineskip 13pt
\nopagenumbers

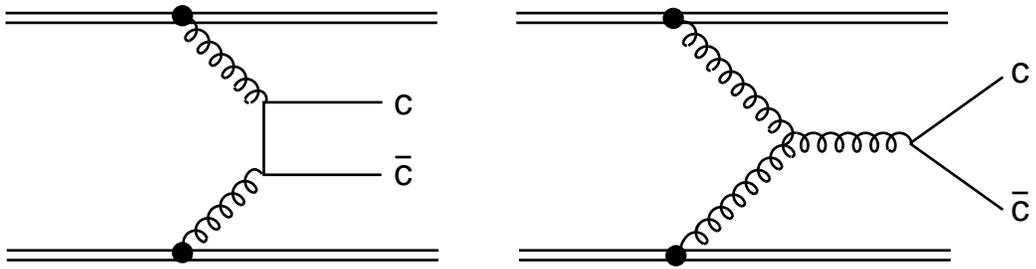
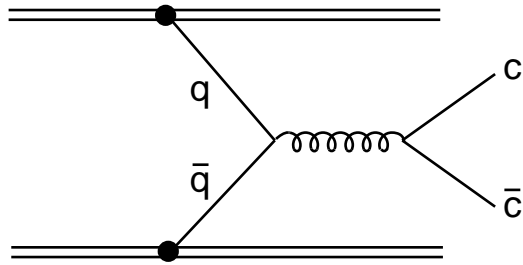
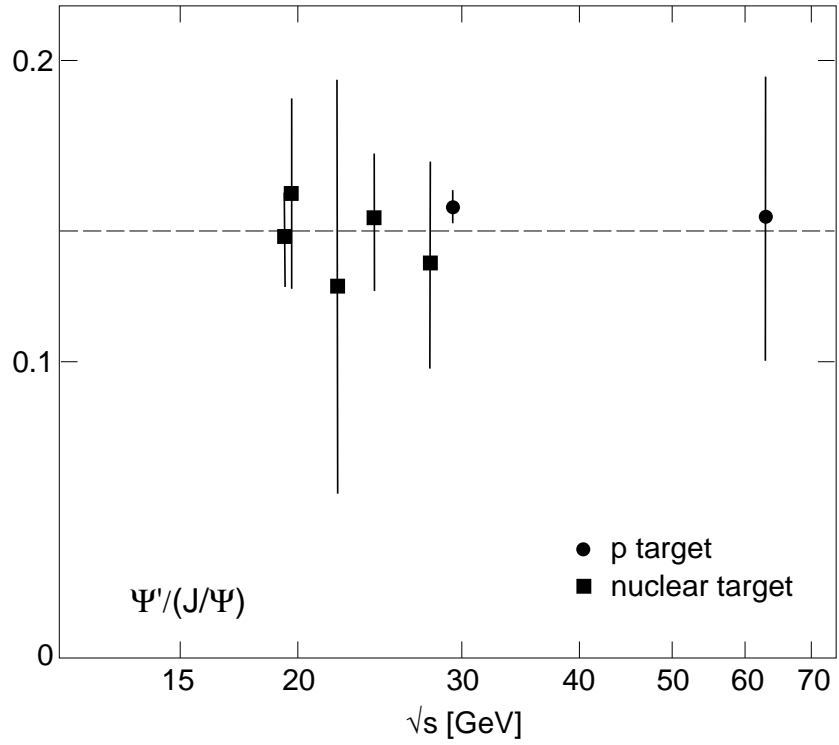
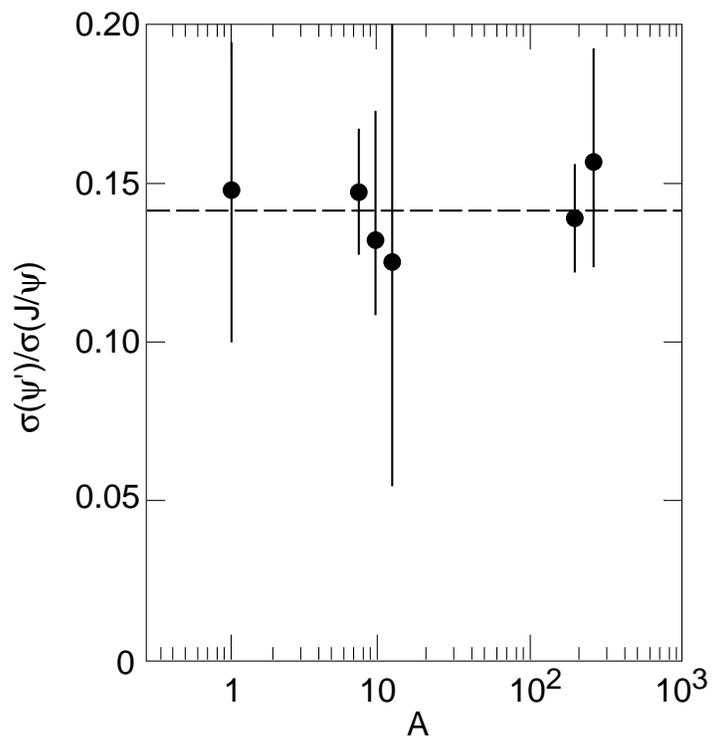
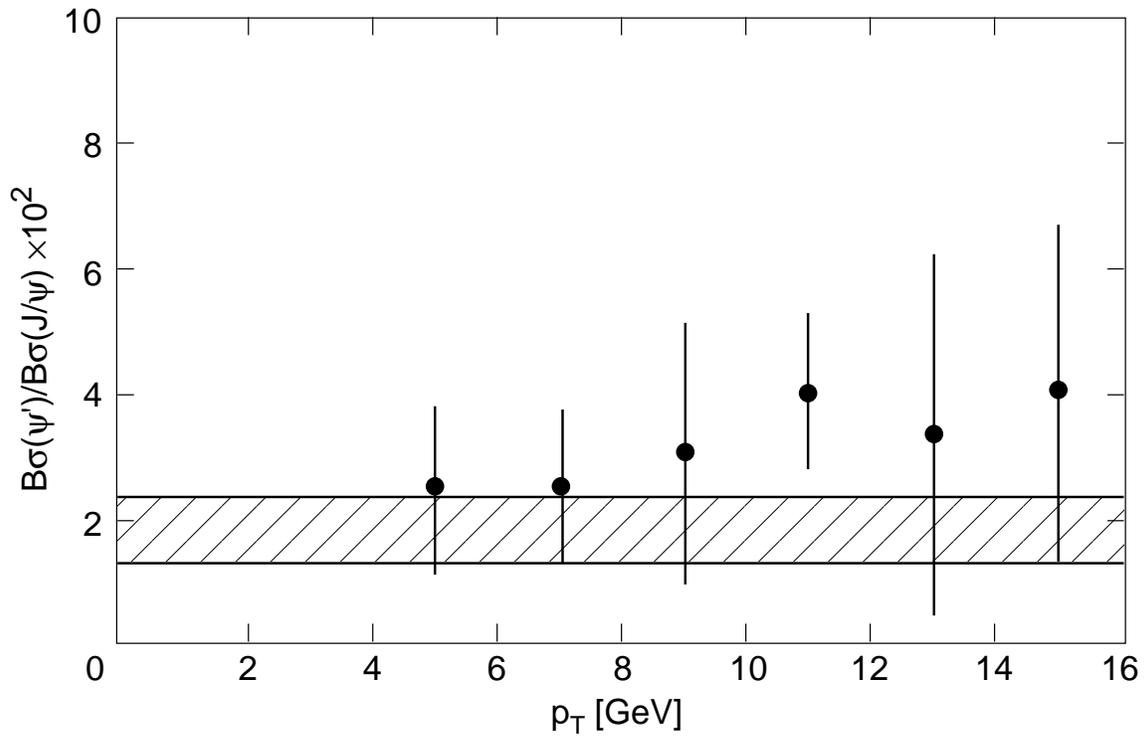
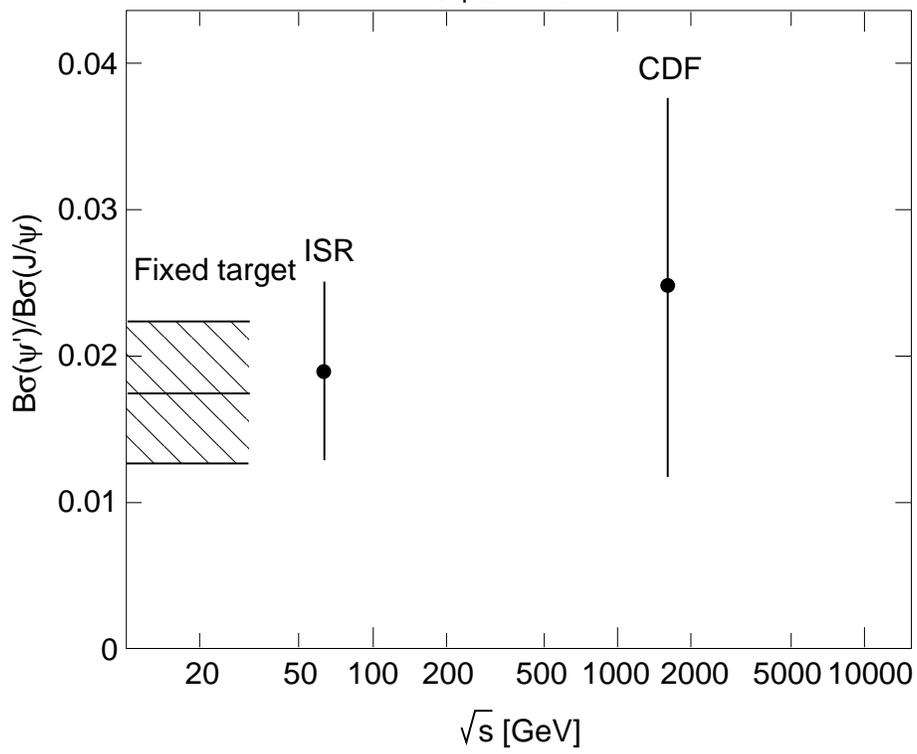
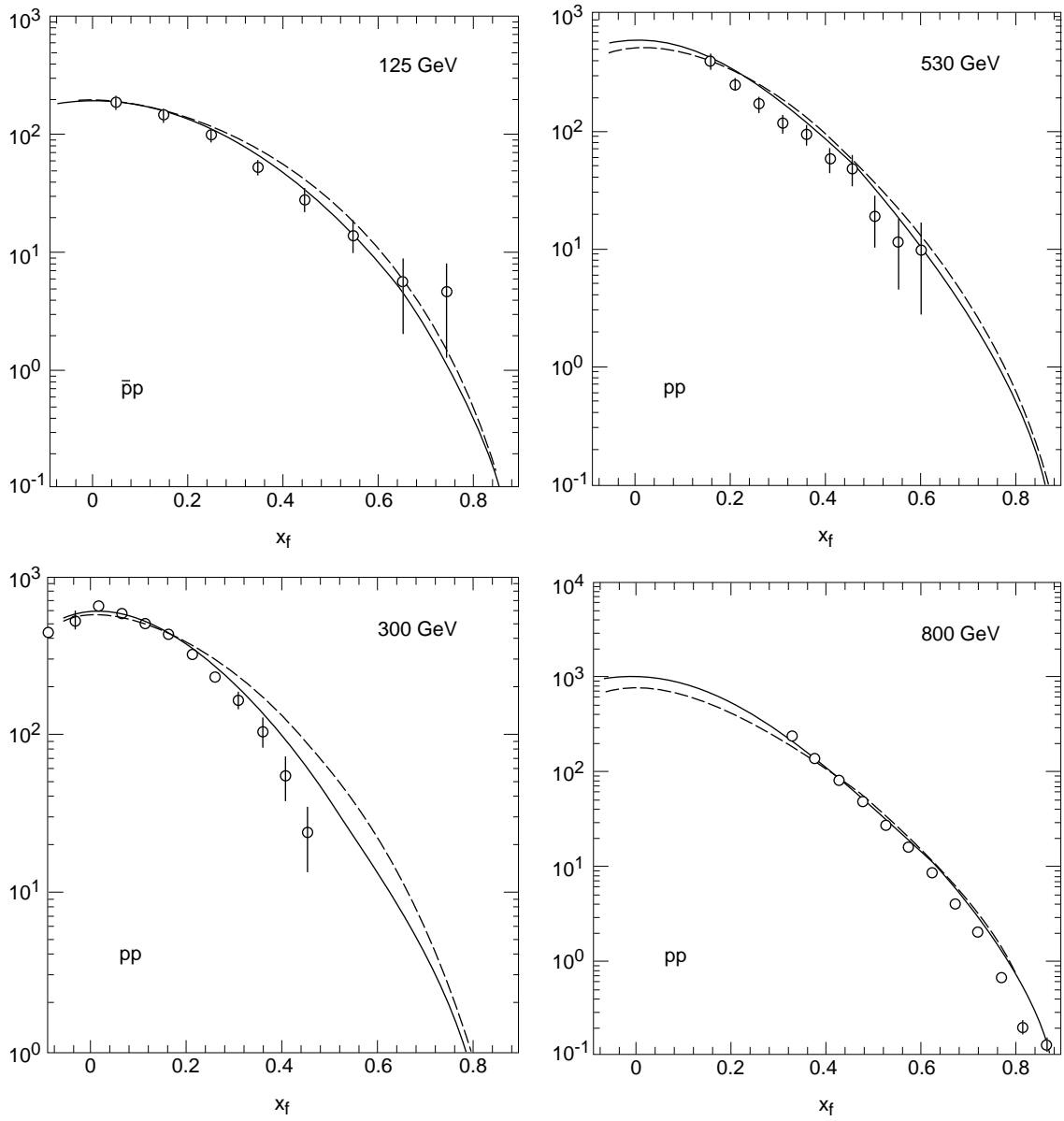
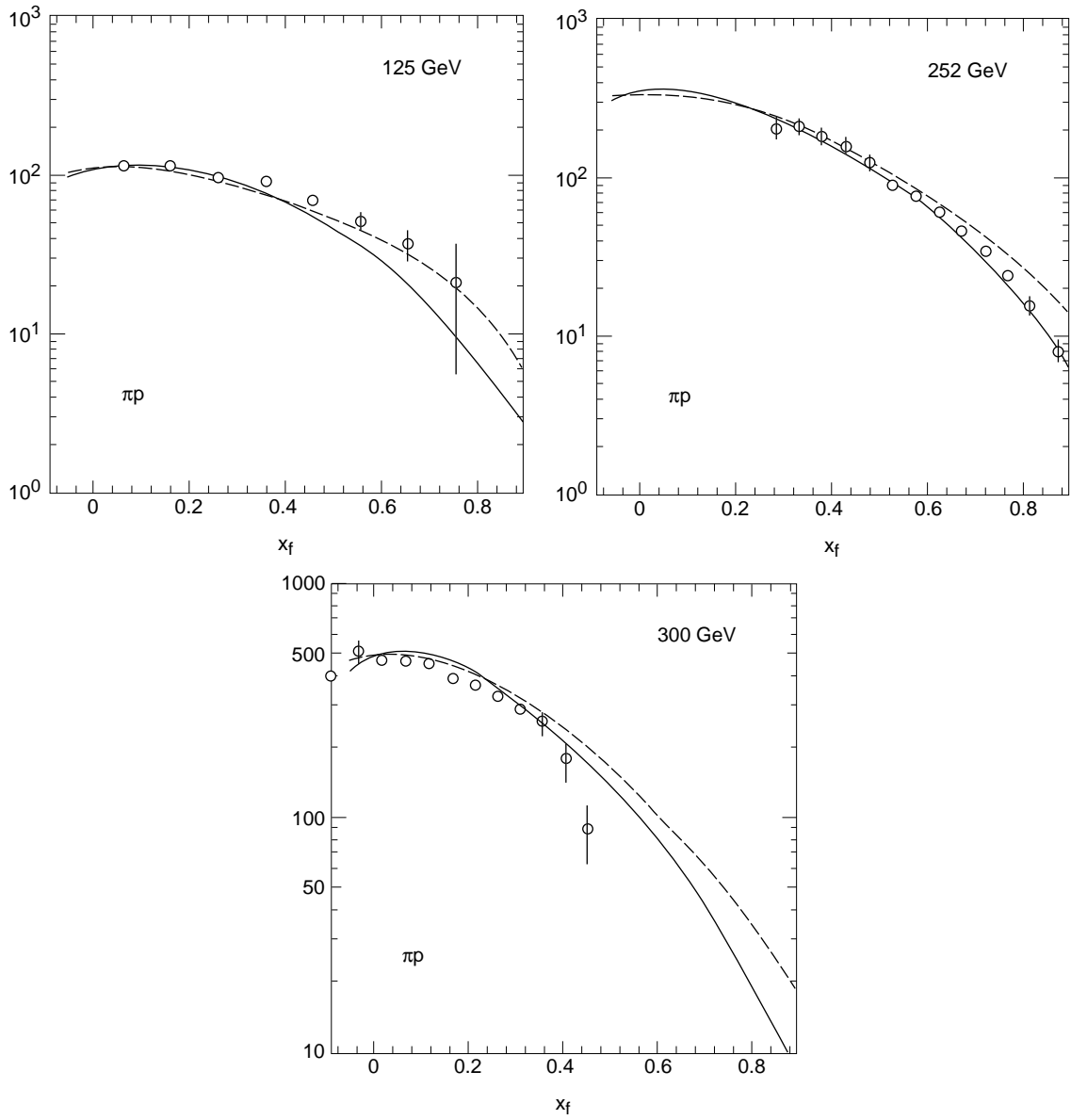
\end